\documentclass[aps, reprint, amsmath, amssymb, floatfix, prl]{revtex4-2}

\usepackage{amsmath,amsthm,amssymb}
\usepackage{graphicx}
\graphicspath{{figs}}

\usepackage{dcolumn}
\usepackage{bm}
\usepackage{color}
\usepackage{epsfig}
\usepackage{multirow}
\usepackage{mathrsfs}
\usepackage{hyperref}
\usepackage{cleveref}
\usepackage{epstopdf}
\usepackage{subfigure}
\usepackage{autobreak}

\usepackage[linesnumbered,ruled,vlined]{algorithm2e}

\newcommand{\V}[1]{\boldsymbol{#1}} 
\newcommand{\M}[1]{\boldsymbol{#1}} 
\newcommand{\abs}[1]{\left|#1\right|} 

\usepackage{xcolor}

\begin{document}

\title{Branch-and-Bound Tensor Networks for Exact Ground-State Characterization}
\author{Yi-Jia Wang$^{1,2}$}\thanks{These authors contributed equally to this work.}
\author{Xuanzhao Gao$^{3}$}\thanks{These authors contributed equally to this work.}
\author{Pan Zhang$^{1,2,4,7}$}\email{panzhang@itp.ac.cn}
\author{Feng Pan$^{5}$}\email{feng\_pan@sutd.edu.sg}
\author{Jin-Guo Liu$^{6}$}\email{jinguoliu@hkust-gz.edu.cn}

\affiliation{$^{1}$CAS Key Laboratory for Theoretical Physics, Institute of Theoretical Physics, Chinese Academy of Sciences, Beijing 100190, China.}
\affiliation{$^{2}$School of Physical Sciences, University of Chinese Academy of Sciences, Beijing 100049, China.}
\affiliation{$^{3}$Center for Computational Mathematics, Flatiron Institute, Simons Foundation, New York, NY 10010, USA.}
\affiliation{$^{4}$School of Fundamental Physics and Mathematical Sciences,\- Hangzhou Institute for Advanced Study, UCAS, Hangzhou 310024, China}
\affiliation{$^{5}$Science, Mathematics and Technology Cluster, Singapore University
 of Technology and Design, 8 Somapah Road, 487372 Singapore}
\affiliation{$^{6}$Hong Kong University of Science and Technology (Guangzhou), Guangzhou 511453, China}
\affiliation{$^7$Beijing Academy of Quantum Information Sciences, Beijing 100193, China}

\date{\today}

\begin{abstract}
    Characterizing the ground-state properties of disordered systems, such as spin glasses and combinatorial optimization problems, is fundamental to science and engineering.
    However, computing exact ground states and counting their degeneracies are generally NP-hard and \#P-hard problems, respectively, posing a formidable challenge for exact algorithms.
    Recently, Tensor Networks methods, which utilize high-dimensional linear algebra and achieve massive hardware parallelization, have emerged as a rapidly developing paradigm for efficiently solving these tasks.
    Despite their success, these methods are fundamentally constrained by the exponential growth of space complexity, which severely limits their scalability.
    To address this bottleneck, we introduce the Branch-and-Bound Tensor Network (BBTN) method, which seamlessly integrates the adaptive search framework of branch-and-bound with the efficient contraction of tropical tensor networks, significantly extending the reach of exact algorithms. 
    We show that BBTN significantly surpasses existing state-of-the-art solvers, setting new benchmarks for exact computation. 
    It pushes the boundaries of tractability to previously unreachable scales, enabling exact ground-state counting for $\pm J$ spin glasses up to $64 \times 64$ and solving Maximum Independent Set problems on King's subgraphs up to $100 \times 100$. For hard instances, BBTN dramatically reduces the computational cost of standard Tropical Tensor Networks, compressing years of runtime into minutes. 
    Furthermore, it outperforms leading integer-programming solvers by over 30$\times$, establishing a versatile and scalable framework for solving hard problems in statistical physics and combinatorial optimization.
\end{abstract}



\maketitle
Disordered systems, encompassing both spin glasses~\cite{binder1986spin, fischer1993spin} and combinatorial optimization problems~\cite{du1998handbook}, offer critical theoretical insights for understanding complex energy landscapes~\cite{zhou2009energy, rodrigues1994spin, ros2023high}, frustration~\cite{toulouse1987theory, kirkpatrick1977frustration}, and glassy dynamics~\cite{cugliandolo2002dynamics}. They are also fundamental to a wide array of applications across science and engineering, ranging from probabilistic inference~\cite{koller2009probabilistic} to quantum information science~\cite{cao2025exact,dutta2015quantum}. 
In statistical physics, studying the ground-state properties of disordered systems elucidates the nature of low-lying excitations and the theory of replica symmetry breaking~\cite{sherrington1975solvable, auffinger2020sk,parisi1979infinite, mezard1987spin}. Similarly, in combinatorial tasks like the maximum independent set (MIS) problem~\cite{du1998handbook}, understanding the solution-space properties is crucial for determining computational complexity~\cite{gamarnik2022algorithms,bresler2022algorithmic}, identifying theoretical phase transitions~\cite{krzakala2007gibbs, daude2008pairs}, and defining the fundamental performance limits of practical solvers~\cite{du1998handbook, Moore2011}.

However, finding exact ground states and counting their degeneracies are NP-hard and \#P-hard, respectively~\cite{barahona1982computational, Moore2011}. 
Traditional exact solvers, such as branch-and-cut~\cite{de1995exact, Andrist2023, Achterberg2009} and branch-and-bound (B\&B) algorithms~\cite{Fomin2006, li2005exploiting,abrame2014ahmaxsat, akiba2016branch}, only provide optimal solutions without counting degeneracies, and rely on task parallelism. Although counting-oriented B\&B methods~\cite{de1995exact, hartwig1984recursive} can enumerate all optimal configurations, they are limited by similar task-based architectural constraints.
Consequently, they all fail to leverage the massive data parallelism of modern hardware, such as GPUs, and are thus computationally intractable for large-scale problems.
Approximate methods like Monte Carlo~\cite{rubinstein2016simulation} or simulated annealing~\cite{kirkpatrick1983optimization} provide no exactness guarantees and often fail in rugged landscapes.
Tensor Network (TN) methods~\cite{markov2008simulating, Schollwock2011, liu2021tropical, Liu2023} have emerged as a powerful exact framework for tackling core challenges in statistical physics and combinatorial optimization. By exploiting global tensor network structure contraction and massive GPU parallelism~\cite{PhysRevResearch.6.033261, Ebadi2022, Roa-Villescas2023}, TN offers significant speedups over traditional serial algorithms. Nevertheless, their scalability faces a fundamental memory bottleneck, as space complexity scales exponentially with the contraction treewidth~\cite{markov2008simulating}. Although the slicing technique~\cite{gray2021hyper} can alleviate memory pressure, it does so at the cost of a prohibitive, exponential time overhead.

In this work, we introduce the branch-and-bound tensor network (BBTN) method, which integrates branch-and-bound search~\cite{Fomin2006, li2005exploiting, abrame2014ahmaxsat, argelich2018clause, eppstein2007traveling, carpaneto1980some} with tropical tensor network contraction to simultaneously reduce space and time complexity. 
By adaptively branching on variables and aggressively pruning configurations using local and global bounds~\cite{Moore2011, Achterberg2007thesis, morrison2016branch, Gao2024}, BBTN constructs unbalanced but dramatically smaller search trees. We evaluate BBTN on two paradigmatic benchmarks: the $\pm J$ Ising spin glass on 2D and 3D lattices, a canonical model for frustration and disorder~\cite{cipra1987introduction, godoy2020ising}, and MIS problems on King's subgraphs, which serve as a critical testbed for neutral-atom quantum processors~\cite{Pichler2018, Ebadi2022, Kim2022}. 
These choices allow us to demonstrate BBTN's efficacy across both physics-inspired and hardware-relevant optimization landscapes. 
Across these benchmarks, BBTN substantially outperforms tensor-network slicing, branch-and-bound methods~\cite{xiao2013,Xiao2017,akiba2016branch}, and state-of-the-art open-source integer-programming solvers~\cite{Achterberg2009}, extending the frontier of classical exact algorithms for spin glass and combinatorial optimization problems relevant to current quantum hardware.
\begin{figure}[htbp]
    \centering
    \includegraphics[width=0.99\linewidth, trim={0.1cm 0.1cm 0.0cm 0.1cm}, clip]{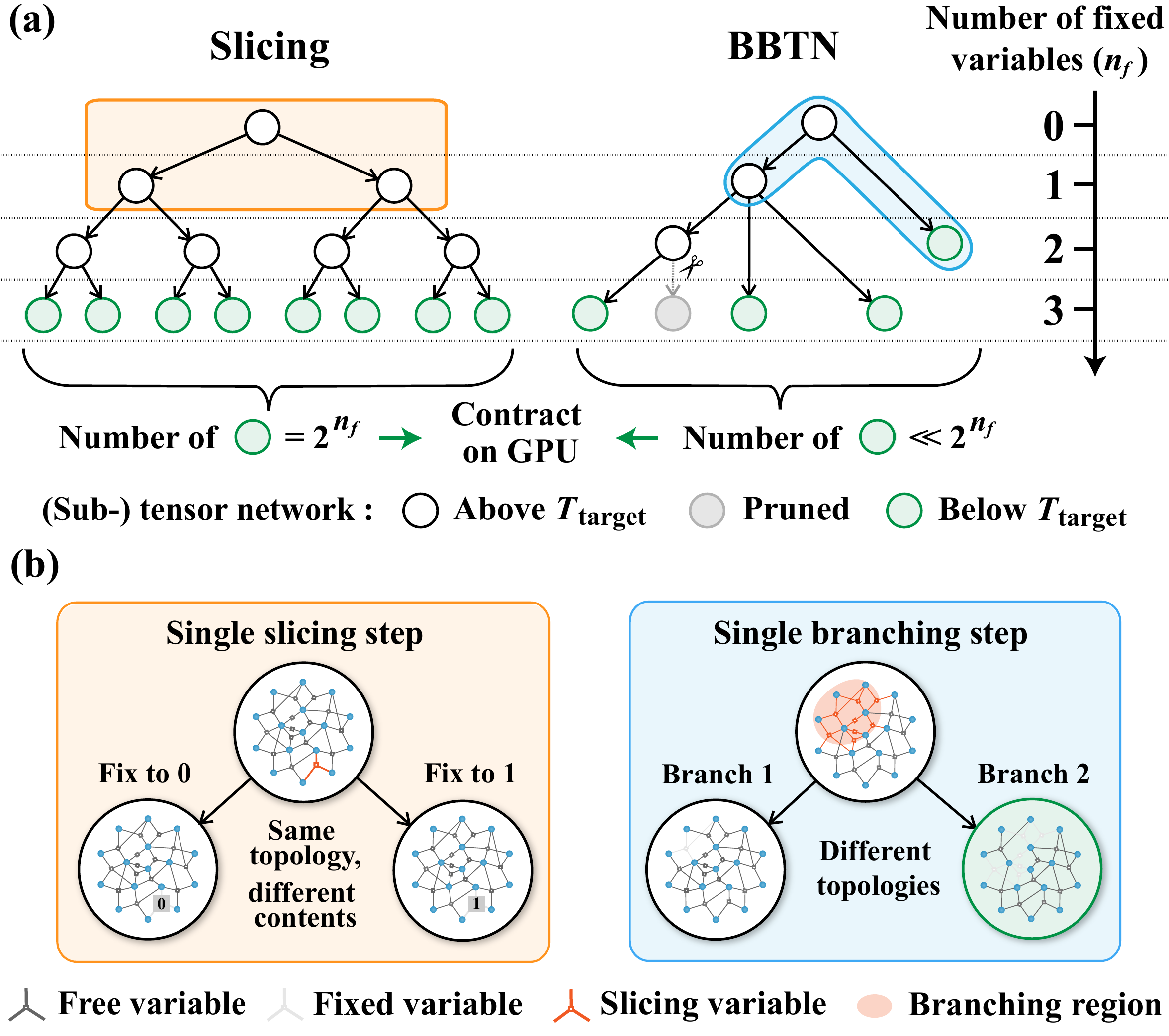}
    \caption{
        Schematic comparison of slicing and branch-and-bound tensor network (BBTN) methods.
        (a) Tensor network decomposition process. Slicing (left) exhaustively fixes a precomputed set of $n_f$ variables, generating $2^{n_f}$ leaf sub-TNs satisfying memory constraints (green nodes). BBTN (right) employs a branch-and-bound framework with 
        pruning and online branching, producing an unbalanced tree with substantially fewer leaf sub-TNs. Dashed arrows show sub-TNs pruned before branching and thus never generated.
        (b) Single-step comparison. Slicing (left) fixes one variable per step, generating two sub-TNs. BBTN (right) examines multiple variables, prunes infeasible or suboptimal configurations, and generates an adaptive number of branches with different subsets of fixed variables.
    }
    \label{fig:bbtn} 
\end{figure}

\textit{Tropical Tensor Networks (TN).}--
We consider the spin glass model defined on a graph $G=(V,E)$ with energy function
\begin{equation}
H(\V{\sigma}) = \sum_{(i,j) \in E} -J_{ij} \sigma_i \sigma_j + \sum_{i \in V} -h_i \sigma_i,
\end{equation}
where $\sigma_i \in \{-1, +1\}$ encodes the $i$-th spin, $J_{ij}$ and $h_i$ denote the coupling constants and external fields, respectively.
Following the graph structure of $G$, we place a matrix $\M{B}^{ij} = \begin{pmatrix} -J_{ij} & J_{ij} \\
J_{ij} & -J_{ij} \end{pmatrix}$ 
on each edge $(i,j)$ and a vector $\V{w}^i = (h_i, -h_i)^T$ on each vertex $i$, with bond index $s_i$ encoding the value of $\sigma_i$. Upon these definitions, the ground state energy $H^*$ can be expressed as a tensor network~\cite{liu2021tropical,Liu2023} using the min-sum tropical algebra with $x \oplus y = \min(x,y)$ and $x \odot y=x+y$:
\begin{equation}\label{eq:tropical_algebra}
    H^* = \bigoplus_{\V{s} \in \{0, 1\}^{\abs{V}}} \left[ \left(\bigodot_{(i,j) \in E} B^{ij}_{s_i s_j}\right) \odot \left(\bigodot_{i \in V} w^i_{s_i} \right)\right].
\end{equation} 
Moreover, since tensor networks with real-number entries can be used to compute the number of solutions~\cite{kourtis2019fast}, we combine them with tropical tensor networks to compute ground-state degeneracies~\cite{liu2021tropical}. 

For the maximum weighted independent set (MWIS) problem, this formalism extends directly; hard constraints can be represented by the tropical number $\infty$ for optimization problems and by the tuple $(\infty, 0)$ for counting problems. 
The size of MWIS on graph $G$ follows the same structure as Eq.~\ref{eq:tropical_algebra}, with the independence constraint $\M{B} = \begin{pmatrix}
    0 & 0 \\
    0 & \infty
\end{pmatrix}$ and vertex tensor $\V{w}^i = (0 \quad -w_i)^T$ encoding the vertex weights. 
Analogously, the counting of the MWIS can be treated in a similar manner.

The tensor networks under the various algebras above are collectively denoted by $\mathcal{T}(G)$, with the contraction result $\texttt{contract}(\mathcal{T}(G))$. 
The computational cost is measured along two dimensions: time complexity (total floating-point operations) and space complexity (memory for the largest intermediate tensor), both of which scale exponentially with the treewidth of the tensor network's line graph~\cite{markov2008simulating}, with the latter, in particular, creating a hard bottleneck.

\textit{TN with Slicing.}--
A standard mitigation technique for this bottleneck is \textit{slicing}, which decomposes the original problem into subproblems by iterating over configurations of a selected variable subset:
\begin{equation}
    V_f := \bigoplus_{\V{x} \in \{0, 1\}^{\abs{V_f}}} \left( \texttt{contract}(\mathcal{T}(G) \mid_{\V{s}_{V_f}=\V{x}}) \right)\;,
\end{equation}
where $\mathcal{T}(G) \mid_{\V{s}_{V_f}=\V{x}}$ denotes the reduced TN with variables in $V_f$ fixed to configuration $\V{x}$ and space complexity below the threshold. 
The subset $V_f$ is precomputed by algorithms such as dynamic slicing~\cite{chen2018classical} to maximize space reduction while minimizing $\abs{V_f}$. 
However, as shown in the left panel of Fig.~\ref{fig:bbtn}, each additional fixed variable doubles the number of sub-TNs with the same topology but different contents, requiring the contraction to be performed $2^{\abs{V_f}}$ times and incurring exponential time overhead. 
This trade-off renders slicing impractical except for very small slicing sets.

\textit{Branch-and-Bound Tensor Network.}--
To overcome this bottleneck, we propose BBTN, which integrates the Branch-and-Bound (B\&B) framework~\cite{Gao2024} into tensor network contraction. 
As illustrated in the right panel of Fig.~\ref{fig:bbtn}, BBTN decomposes a tensor network into multiple smaller sub-TNs by fixing a subset of variables just like the slicing technique.
However, BBTN introduces two key distinctions. 
The first one is \emph{pruning}, while slicing exhaustively processes all sub-TNs, BBTN {prunes} infeasible or suboptimal sub-TNs by evaluating local energy contributions over a subset of variables $R \subseteq V$ (referred to as the branching region). 
The second one is \emph{branching}: BBTN generalizes slicing by allowing for adaptive variable assignment, where different subsets of $R$ are fixed across different branches. Consequently, BBTN constructs a non-uniform search tree that is significantly smaller than that of slicing, potentially offering an exponential reduction in contraction cost.

In detail, the pruning step systematically exploits solution space sparsity using global and local bounds derived from the B\&B framework. 
Global pruning maintains an upper bound of the ground state based on the lowest energy found so far; any branch with a lower bound exceeding this threshold is discarded. 
Local pruning, conversely, operates on the sub-TN $\mathcal{T}_R$ and relies on two criteria: the boundary-equivalence principle, which retains only the optimal configuration for each unique boundary state, and logical inference rules~\cite{sm} that eliminate provably suboptimal configurations. 
Together, these mechanisms yield a minimal set of feasible local configurations, $\mathcal{F}(R) \subseteq \{0,1\}^{|R|}$, where every variable in $R$ is assigned a binary value.

After pruning, the feasible set $\mathcal{F}(R)$ represents the valid configurations within the search space. A direct partition of $\mathcal{F}(R)$ into uniform subspaces yields an intractable number of branches~\cite{sm}. 
Instead, BBTN employs a non-uniform branching strategy, defining a set of branches $\mathcal{D}(R) \subseteq \left\{ (R_f,\V{x}) \mid R_f \subseteq R, \; \V{x}\in \{0,1\}^{|R_f|} \right\}$ where the subset of fixed variables $R_f \subseteq R$ varies. 
For exact counting, this decomposition must be complete, covering all feasible configurations in $\mathcal{F}(R)$, and orthogonal, preventing double-counting.
The final result is obtained by summing the contractions of each branch. 
To optimize $\mathcal{D}(R)$, we employ the Online Branching (OB) scheme~\cite{Gao2024} to minimize the branching factor $\gamma$. This approach builds upon the common assumption in conventional B\&B frameworks that time complexity is exponential in $\gamma$ and the problem size $\rho$, i.e., $O(\gamma^{\rho})$. 
Crucially, to address the space complexity bottleneck, we define the problem size $\rho$ as the cumulative memory overflow:
\begin{equation}
\rho(\mathcal{T}) = \sum_{\mathbf{T}} \max \left( 0, ~\log_2 |\mathbf{T}|- \log_2 T_{\rm target} \right),
\end{equation}
where $|\mathbf{T}|$ is the tensor size and $T_{\rm target}$ is the memory threshold. The optimal branching rule is found by minimizing $\gamma$ subject to the characteristic equation $\gamma^{\rho(\mathcal{T})} = \sum_{i} \gamma^{\rho(\mathcal{T}_i)}$, thereby systematically reducing the number of sub-TNs.

\textit{Applications.}--
We evaluate BBTN on two classes of instances: the $\pm J$ spin glasses, a canonical model that has been extensively studied as a fundamental benchmark in spin glass theory, and the MIS problem on King's subgraphs (KSG), which has recently played an important role as a key benchmark for assessing neutral-atom quantum processors~\cite{Ebadi2022}. 
For the spin glass model, we set the coupling constants $J$ to be randomly chosen as $+1$ or $-1$, and fix the external field at $0.5$, which admits no known analytical solution. 
The KSG instances comprise random KSG (RKSG) with filling rate $0.8$, typical for experimental hardware~\cite{Manetsch2024}, and mapped KSG (MKSG) that encode integer factorization problems~\cite{Nguyen2023}, where an $n$-bit composite integer $m=pq$ is factored into prime factors $p$ and $q$.

\begin{figure}[htbp]
    \centering
    \includegraphics[width=0.95\linewidth,trim={2.5cm 0.0cm 2.5cm 0.0cm}, clip]{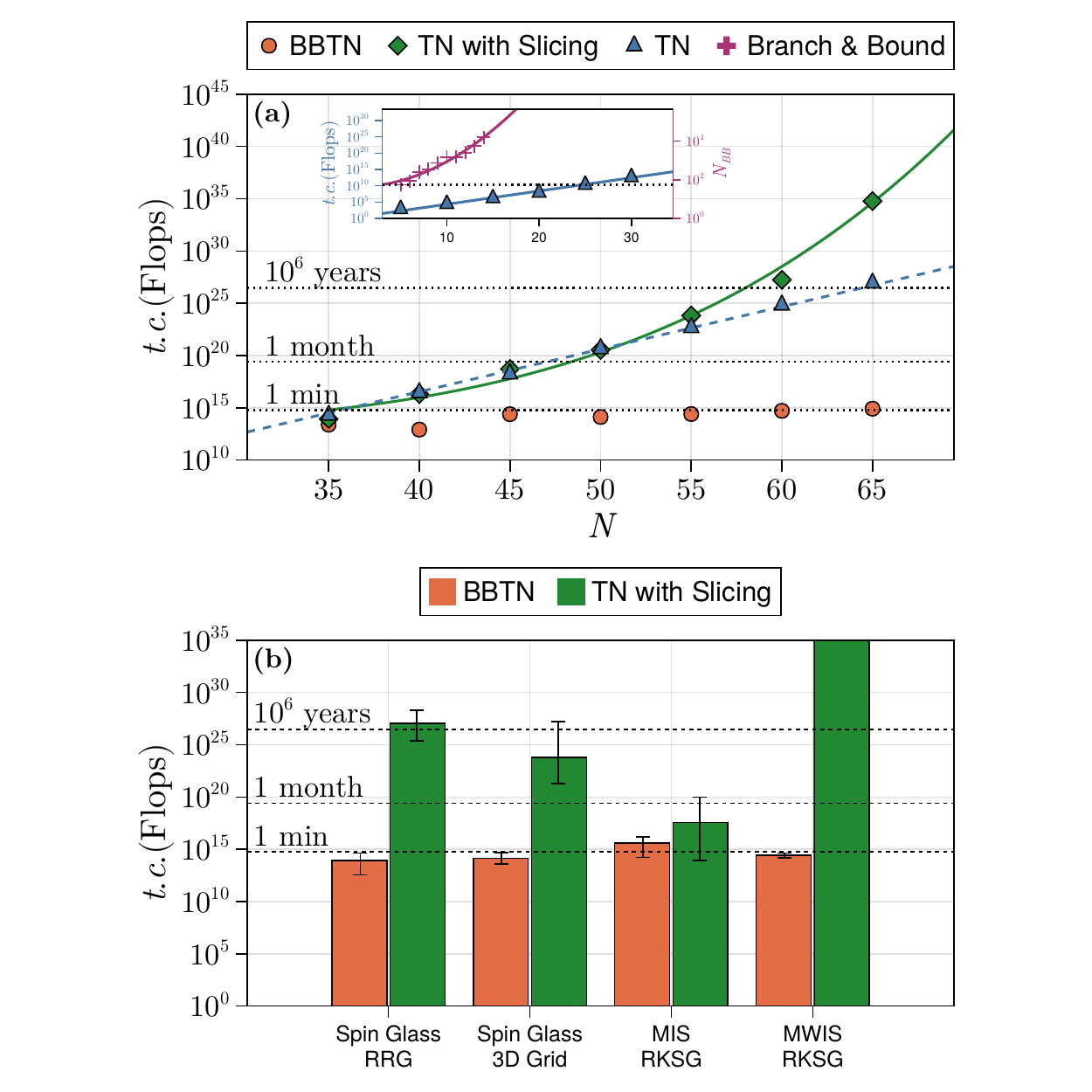}
    \caption{
        (\textit{Top}) Performance comparison of BBTN, tropical-TN with and without slicing, and vanilla branch-and-bound methods (shown in the inset) on ground-state counting for spin glasses on $N\times N$ 2D lattices ($J=\pm1,h=0.5$). 
        Averages for the solid dots are computed over all 10 instances. 
        Tensor network runtimes are calibrated from theoretical time complexity and peak FLOPS of NVIDIA A100 GPU. The time reference line in the inset indicates that both methods have actual runtimes on the order of seconds.
        (\textit{Bottom}) Performance of BBTN and slicing for ground-state counting on four problem classes: spin glasses ($J=\pm1,h=0.5$) on random regular graphs ($n=600,d=3$) and on an $8 \times 8 \times 8$ 3D lattice, MIS on RKSG ($n=60$, filling$=0.8$), and MWIS on RKSG ($n=60$, filling$=0.8$, with weights uniformly sampled from integers 1 to 10).  
        Bars denote mean runtimes over 10 instances per class, with error bars spanning the min--max across instances. 
    }
    \label{fig:ground_counting}
\end{figure}

For both problem classes, we first consider counting ground-state solutions. 
Since transformations that break ground-state degeneracy are inapplicable, the only viable classical algorithm is branch-and-bound~\cite{hartwig1984recursive}. 
We therefore employ two baselines for comparison: the tropical tensor network with slicing, whose slicing set is generated by the dynamic slicing algorithm~\cite{Liu2023}, and a branch-and-bound algorithm that uses the same bounding strategy as BBTN. 
To ensure a fair comparison, we constrain both slicing and BBTN to run on a single GPU, fixing their space-complexity target at $2^{31}$~\cite{sm}. 
Additionally, we include the plain tropical TN method as an exponential baseline, which exceeds the memory target at $N=31$.
As shown in Fig.~\ref{fig:ground_counting} (\textit{Top}), for the $\pm J$ spin glass model on $N \times N$ 2D lattices, both B\&B (shown in the inset) and slicing exhibit superexponential time complexity growth: B\&B can only handle instances with $N \leq 15$, which are easily computed by tropical-TN, while slicing requires an estimated $10{,}000+$ years to reach $N=60$.
In contrast, BBTN's complexity increases much more slowly as $N$ increases, and solves instances with $N$ up to $65$.
Fig.~\ref{fig:ground_counting} (\textit{Bottom}) further shows that BBTN consistently outperforms slicing across diverse graph topologies for spin glasses, as well as for MIS and MWIS problems, demonstrating its broad applicability.

\begin{figure}[htbp]
    \centering
    \includegraphics[width=0.93\linewidth]{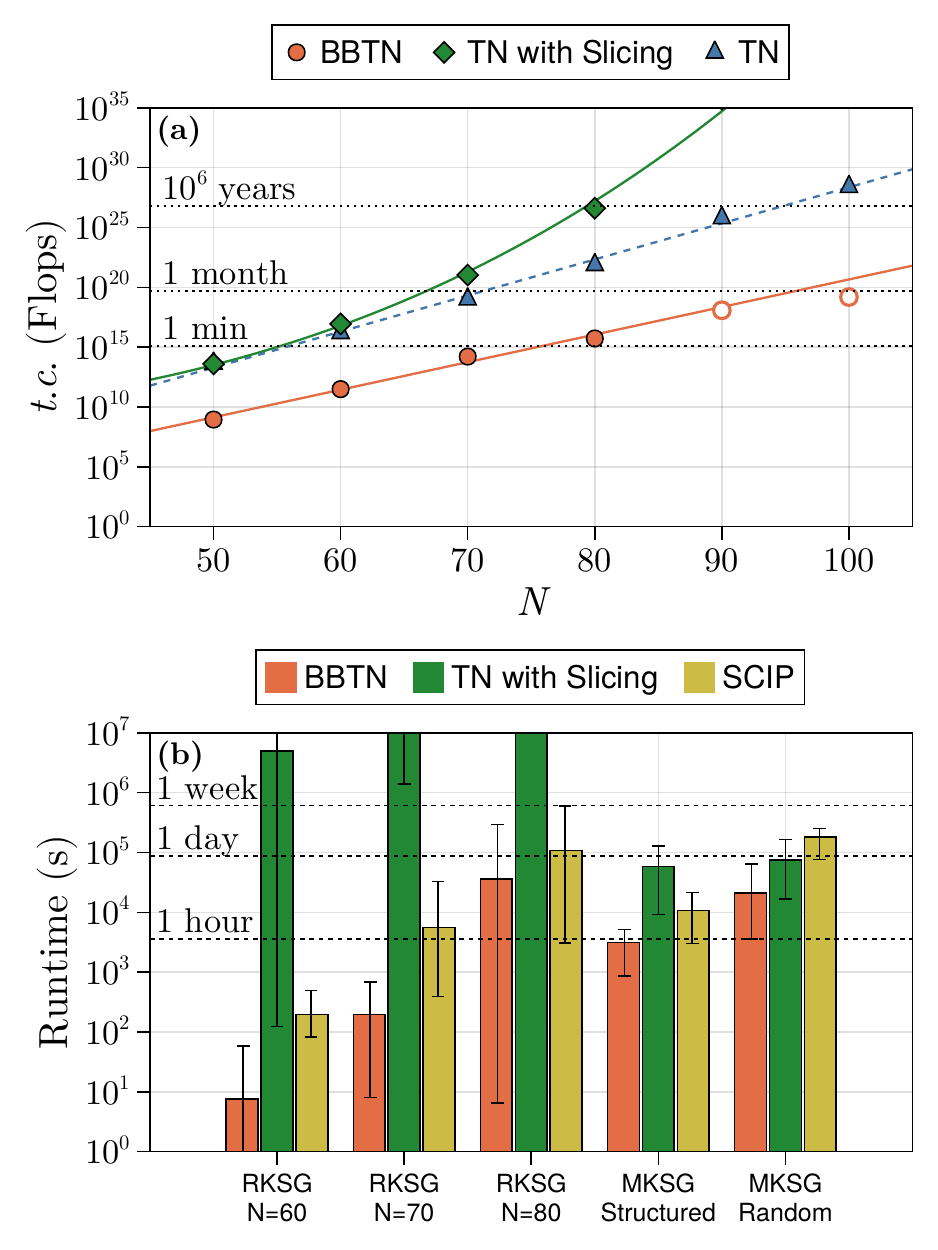}
    \caption{(\textit{Top}) Performance comparison of BBTN, tropical-TN with and without slicing on optimal solution computation for MIS problems on $N\times N$ RKSG (filling 0.8). 
    Averages for the solid dots are computed over all 10 instances. For BBTN, hollow dots show averages over only the instances solvable in time (4/10 for $N=90$ and 2/10 for $N=100$), where the scaling curves are fits to the solid dots.
       (\textit{Bottom}) Runtime comparison of BBTN, slicing, and SCIP. We benchmark on two problem classes: MIS on RKSG and MWIS on MKSG. 
        Specifically, for slicing, we contract a single slice to estimate total runtime, as all slices are identical.
    }
    \label{fig:ground_state}
\end{figure}

For ground-state search, BBTN incorporates advanced branch-and-cut bounds, proving highly efficient for MIS and MWIS problems. We benchmark performance against slicing and SCIP~\cite{Andrist2023}, while omitting traditional graph-based B\&B algorithms due to their inability to scale to these system sizes~\cite{xiao2013, Xiao2017, hespe2020wegotyoucovered, sm}.

We begin with the random KSG with filling rate $0.8$ and unit weight, and a comparison of the time complexity of BBTN, tropical-TN, and slicing is shown in Fig.~\ref{fig:ground_state} (\textit{Top}).
As in the ground-state degeneracy calculations, the tropical tensor-network method quickly exceeds the memory limit and becomes infeasible, and slicing suffers from exponential growth in runtime, rendering it impractical for $N>70$. 
In contrast, BBTN maintains both time and memory within manageable limits, successfully solving systems up to $N=100$ (corresponding to 8000 vertices).

A wall-clock runtime comparison of BBTN, slicing, and SCIP is presented in Fig.~\ref{fig:ground_state} (\textit{Bottom}) for both MIS and MWIS problems.
The initial three bars depict results for RKSG instances with $N=60,70,80$. 
Slicing again suffers from exponential runtime growth and is estimated to take months or even years to complete, whereas BBTN consistently finds solutions within seconds, achieving a 30× speedup over SCIP.
The subsequent two bars demonstrate BBTN's robustness across different optimization landscapes for MWIS on MKSG encoding integer factorization problems, which possess only two degenerate optima, making them highly sensitive to landscape structure compared to high-degeneracy RKSG.
We investigate two prime ensembles: \emph{random} 16-bit primes with only the most significant bit fixed to 1, and \emph{structured} primes where the nine most significant bits are all 1.
The random ensemble generates rugged landscapes that challenge local-search algorithms~\cite{sm}, while the structured ensemble yields an intrinsically structured landscape emerging from the logarithmic complexity of partial factoring~\cite{coppersmith1996finding}.
BBTN substantially outperforms both slicing and SCIP across both landscape types, achieving nearly 10× speedup over SCIP on rugged landscapes and nearly 20× speedup over slicing on structured landscapes.

\textit{Discussion.}-- BBTN's success on spin-glass and MIS problems—prototypical and specific NP-hard models, respectively—demonstrates its immediate applicability to diverse combinatorial optimization problems, including dominating set, graph coloring, and $k$-SAT. Beyond optimization, the framework's ability to handle sparse tensor network contractions allows it to address fundamental tasks such as quantum circuit simulation and quantum error correction. By extending the underlying algebra to generic semirings, BBTN can also characterize complex solution-space properties, such as solution configurations and $k$-largest spectra. Furthermore, relaxing exactness constraints naturally extends the framework to high-fidelity approximate contractions.

The paradigm established by BBTN can be synergistically integrated with AI-driven and mathematical optimization frameworks to achieve even greater efficacy. Specifically, the current reliance on step-wise optimization of local heuristic measures—which is computationally intensive and lacks global foresight—could be replaced by deep reinforcement learning and Monte Carlo Tree Search, which would substitute myopic decision-making with a strategic, global policy. Additionally, integrating advanced techniques from the mathematical programming community, such as branch-and-cut and branch-and-price, offers sophisticated solution-space transformations that complement tensor network contractions. 

\begin{acknowledgments}
\textit{Acknowledgements}-- 
This work is partially supported by the National Natural Science Foundation of China under grant nos. 12325501, 12404568, 12047503, and 12247104, the National Key R\&D Program of China (Grant No.~2024YFB4504004), and the Guangzhou Municipal Science and Technology Project (No. 2024A03J0607). FP acknowledges the support from the Ministry of Education Singapore under grant No. SKI 2021\_07\_03 and National Quantum Computing Hub translational fund.
The authors also thank Zisong Shen, Cheng Ye, Yan Mi, Xiwei Pan for helpful discussions on the manuscript.

The Julia~\cite{julia-2017} implementation of our algorithm is available at \cite{BBTN}.
For tropical tensor networks and dynamic slicing, we use \texttt{GenericTensorNetworks.jl}~\cite{Liu2023, gao2025programming}.
\texttt{SCIP}~\cite{Achterberg2008, Achterberg2009} is called via \texttt{JuMP.jl}~\cite{Dunning2017, Lubin2023}.
All CPU experiments run on an Intel Xeon Platinum 8268 (2.90\,GHz), and GPU experiments on an NVIDIA A100.
\end{acknowledgments}

\bibliography{refs}

\end{document}